\documentclass[amsmath,amssymb,aps,prb,reprint,superscriptaddress,nofootinbib]{revtex4-2}
\usepackage{lmodern}
\usepackage{amssymb}
\usepackage{amsmath}
\usepackage{bbm}
\usepackage{mathtools}
\usepackage{braket}
\usepackage{color}
\usepackage{enumerate}
\usepackage[colorlinks=true,
            linkcolor=white!40!red,
            citecolor=green!70!blue,
            urlcolor=blue!80!white]{hyperref}
\usepackage{cancel}
\usepackage{tikz}
\usetikzlibrary{calc, patterns, patterns.meta}

\begin{document}

\title{Transport and thermalization under noncommuting localized constraints}

\author{Amir Nesturov}%
\affiliation{%
Department of Physics, Faculty of Mathematics and Physics, University of Ljubljana, Jadranska 19, Ljubljana SI-1000, Slovenia
}
\author{Attila Tak\'acs}%
\affiliation{%
Department of Physics, Faculty of Mathematics and Physics, University of Ljubljana, Jadranska 19, Ljubljana SI-1000, Slovenia
}
\author{Lenart Zadnik}
\email{lenart.zadnik@fmf.uni-lj.si}
\affiliation{%
Department of Physics, Faculty of Mathematics and Physics, University of Ljubljana, Jadranska 19, Ljubljana SI-1000, Slovenia
}

\begin{abstract}
We investigate the effects of spatially localized conserved quantities on transport and thermalization in quantum and classical many-body systems. We consider several related models: the integrable quantum Volterra model, its nonintegrable classical limit, and two cellular automata derived from the latter. Although localized constraints are often associated with lattice gauge theories, our models do not admit such an interpretation. In systems with discrete degrees of freedom, these constraints induce uniform fragmentation of the Hilbert or configuration space. In the quantum Volterra model, we derive an exact mapping of the dynamics within each Hilbert-space fragment onto a quantum chiral Potts chain. We find that energy transport is ballistic, with identical transport coefficients across all fragments. In the nonintegrable classical limit, energy transport is diffusive, and we similarly find the diffusion constant to be independent of whether or not the constraints are fixed (resolved). Finally, in the cellular automata, we show that, despite leaving transport unaffected, localized constraints lead to a strong breakdown of local ergodicity.
\end{abstract}

\maketitle

\section{Introduction}

Thermalization and transport are two of the busiest avenues for probing universal behavior in many-body systems. While generic systems are expected to exhibit chaotic dynamics, local thermalization, and diffusive charge transport, a variety of mechanisms can obstruct this conventional behavior. These include integrability, disorder, and various types of kinetic constraints~\cite{Calabrese_2016,Garrahan_2018,Gopalakrishnan_2020,Sierant_2025}. All of them can give rise to anomalous transport and to different forms of thermalization breakdown~\cite{Turner_2018,Iadecola_2019,Sala_2020,Khemani_2020,Morningstar_2020,Ilievski_2021,Singh_2021,Yang_2022,Zadnik_2023,Brighi_2023,Bhakuni_2025,Maric_2026,Muzzi_2026,Yoshimura_2026}. In quantum many-body systems, one particularly striking mechanism of ergodicity breaking is Hilbert-space fragmentation (see, e.g., Ref.~\cite{Moudgalya_Bernevig_2022} for a review). In a fragmented system, the Hilbert space decomposes into exponentially many dynamically disconnected subspaces. This phenomenon is not exclusive to quantum systems: it can also occur in classical systems with a finite phase space, which can similarly separate into dynamically disconnected components. The individual sectors of the phase space can differ substantially in size and in the dynamics they support. As a consequence, relaxation and transport may depend strongly on the initial state or configuration~\cite{Pancotti_2020,Yang_2020,Bastianello_2022,Bidzhiev_2022,Zadnik_2022,Will_2023,Kasim_2025,Classen-Howes_2025,Kasim_2026,OrlovP_2026}.

The term Hilbert-space fragmentation encompasses rather different situations. In some systems, no single fragment dominates the Hilbert space, while in others a large thermalizing sector coexists with much smaller atypical ones. An extremal example of the former situation is provided by systems with an extensive number of localized constraints. This occurs, for example, in lattice gauge theories, where local Gauss laws divide the Hilbert space into dynamically disconnected sectors~\cite{Smith_2017,Brenes_2018,Ciavarella_2025,Budde_2026}. The existence of such an extremal form of fragmentation already illustrates that fragmentation is, by itself, primarily a statement about the decomposition of the accessible state space. Its dynamical consequences need not be universal, but may depend on the structure and dynamics of the individual fragments. This raises a more fundamental question: which aspects of many-body dynamics follow directly from fragmentation, and which require taking into consideration the specific properties of the fragmented sectors?

In this work, we investigate this question in a particularly simple setting, which we refer to as \emph{uniform fragmentation}. Here, all dynamically disconnected sectors of the Hilbert space have the same size. Since it lacks variation in the fragment size, this setting provides a natural framework for asking whether fragmentation itself implies nontrivial consequences for transport. We consider four related models: the integrable quantum Volterra model~\cite{Volkov_1989,Volkov_1992}, its simplest nonintegrable classical limit, and two cellular automata---one linear and one nonlinear---based on a time discretization of the classical limit. A common feature of these dynamical systems is an extensive set of localized noncommuting constraints, which play a central role in our analysis. In the quantum Volterra model and the cellular automata, where the degrees of freedom are discrete, these constraints induce uniform fragmentation: the Hilbert space and the discrete configuration space, respectively, decompose into exponentially many sectors of equal size. In the continuous classical Hamiltonian limit the constraints instead confine the dynamics to uncountably many invariant submanifolds of the phase space.

First, by resolving a maximal commuting subset of the constraints in the quantum Volterra model, we obtain a direct mapping of the dynamics within each fragment onto a quantum chiral Potts chain. To the best of our knowledge, such a direct mapping has not been reported previously. We then introduce a nonintegrable classical limit of the model, using which we argue that the localized constraints do not form a first-class constraint algebra~\cite{Henneaux_1992}, distinguishing our models from conventional lattice gauge theories. Next, we study how uniform fragmentation manifests itself in transport and ergodic properties. In the quantum Volterra model, energy transport is ballistic, and the transport properties are identical across all fragments. We find an analogous insensitivity to the constraints in a nonintegrable classical limit of the model, where energy transport is diffusive and the same diffusion constant is obtained with the constraints either resolved or unresolved. Finally, using cellular automata that inherit the same structure of localized constraints, and which are defined on a finite and discrete configuration space, we quantify the breakdown of local ergodicity associated with fragmentation. We show that the distance between the probability distribution of configurations along an individual orbit and the microcanonical distribution over the full phase space, both projected onto a local subsystem and averaged over orbits, approaches a nonzero constant. Under a specific assumption, we show that this constant coincides with the distance between the subsystem projections of the microcanonical distributions defined on the full phase space and on an individual fragment. The nonvanishing distance demonstrates that long-time averages of local observables need not agree with their microcanonical expectation values over the full phase space. Combined, our results indicate that, while the disconnected structure of the phase space remains directly visible in the system's ergodic properties, uniform fragmentation need not affect bulk transport.

The rest of the paper is organized as follows. In Section~\ref{sec:Volterra} we introduce the quantum Volterra model, demonstrate the uniform fragmentation of its Hilbert space, and derive the Hamiltonian governing the dynamics within individual fragments. In Section~\ref{sec:not-a-gauge-theory}, we consider the nonintegrable classical limit of the quantum Volterra model and use it to argue that the constraints considered herein do not define a conventional lattice gauge theory. In Section~\ref{sec:energy-transport} we investigate the effect of localized constraints on energy transport in the quantum Volterra model and its nonintegrable classical limit. Finally, in Section~\ref{sec:thermalization} we  introduce two cellular automata with discrete configuration space and investigate the breakdown of local ergodicity due to localized constraints.

\section{Quantum Volterra model}
\label{sec:Volterra}

To illustrate how uniform fragmentation arises from localized constraints, we first focus on the quantum Volterra model~\cite{Volkov_1989,Volkov_1992}. We consider a one-dimensional chain consisting of $2L$ sites, each one equipped with operators $u$ and $v$ satisfying the Weyl algebra
\begin{align}
    u v =q \, v u.
\label{eq:Weyl-relation}
\end{align}
Here, $q$ is a complex parameter on a unit circle. Setting $q=e^{2\pi i/3}$ for simplicity, the operators $u$ and $v$ can be represented as matrices
\begin{align}
    u=
    \begin{pmatrix}
        1 & 0 & 0 \\
        0 & q & 0 \\
        0 & 0 & q^2
    \end{pmatrix},
\qquad
    v=\begin{pmatrix}
        0 & 0 & 1 \\
        1 & 0 & 0 \\
        0 & 1 & 0
    \end{pmatrix}.
    \label{eq:u_v-matrices}
\end{align}
We extend them to act on the full chain as $u_n=\mathbbm{1}_{3^{n-1}}\otimes u\otimes \mathbbm{1}_{3^{2L-n}}$ and $v_n=\mathbbm{1}_{3^{n-1}}\otimes v\otimes \mathbbm{1}_{3^{2L-n}}$, where $\mathbbm{1}_d$ denotes a $d\times d$ identity matrix.\footnote{We will skip the index $d$ whenever the dimension is clear from the context.} The Hamiltonian of the (3-state) quantum Volterra model with periodic boundary conditions $u_0\equiv u_{2L}$ and $v_0\equiv v_{2L}$ is now
\begin{align}
    H_{{\rm V}}=\sum_{n=1}^{2L} h_{n-1,n},\qquad h_{n-1,n}=-\frac{ w_n^{}+w_n^{-1}}{\sqrt{3}},
    \label{eq:q-Volterra-model}
\end{align}
where the two-site operators 
\begin{equation}
w_n=u_{n-1}v_{n-1} u_n v_n^{-1}
\label{w-def}
\end{equation}
obey exchange relations 
\begin{align}
\begin{aligned}
    &w_{n-1} w_n = q^{2} w_n w_{n-1},
    \\
    &[w_m,w_n]=0 \, \text{ for } \, |m-n|>1.
    \label{eq:w-exchange-algebra}
\end{aligned}
\end{align}
This model is Bethe-ansatz solvable and can be generalized to an arbitrary root of unity $q^N=1$~\cite{Volkov_1992}. The generalization requires a modification of the local density $h_{n-1,n}$ and admits an integrable classical limit. The latter is obtained by sending $N\to\infty$ and replacing the commutator with a Poisson bracket. In this limit, the Hamiltonian equations describe the predator-prey sequence~\cite{Volkov_1989,Volkov_1992}, while the exchange relations~\eqref{eq:w-exchange-algebra} reduce to a lattice deformation of the Virasoro algebra~\cite{Faddeev_1986}.

Returning to the case $q^3=1$, all $w_n$ commute with global (on-site) operators
\begin{align}
    U_1=\prod_{n=1}^{2L} u_n,
    \qquad
    U_2=\prod_{n=1}^{L}v^{-1}_{2n-1}v^{}_{2n},
    \label{eq:global-symmetries}
\end{align} 
which therefore correspond to two (on-site) $\mathbb{Z}_3$ symmetries of the quantum Volterra model, i.e., $[H_{\rm V},U_{1,2}]=0$. In addition to the on-site symmetries $U_{1,2}$, consider the ``dual'' operators
\begin{align}
    \tilde{w}^{}_n=u^{}_{n-1}v^{-1}_{n-1}u_nv_n,
\end{align}
obeying exchange algebra similar to the one in Eq.~\eqref{eq:w-exchange-algebra}. As a courtesy of the Weyl relation~\eqref{eq:Weyl-relation}, we have 
\begin{align}
    [w_m,\tilde{w}_n]=0,\quad \forall m,n,
\label{eq:commutant-bracket}
\end{align}
so the dual operators constitute an extensive set $\{\tilde{w}_n\}$ of independent and noncommuting localized constraints for the Hamiltonian~\eqref{eq:q-Volterra-model}. Due to Eq.~\eqref{eq:commutant-bracket}, these constraints commute with the $\mathbb{Z}_3$ symmetries $U_{1,2}$. Indeed, $U_{1,2}$ can be expressed solely in terms of $w$-s by noting that
\begin{align}
    U_1U_2=w_{1}w_{3}\cdots w_{2L-1},
    \quad
    U^{}_1U_2^{-1}=w_{2}w_4\cdots w_{2L},
\end{align}
and $(U_{1,2})^3=\mathbbm{1}$.

In the following, we will show that the extensive number of mutually-noncommuting constraints $\tilde{w}_n$, which commute with all local terms in the Hamiltonian~\eqref{eq:q-Volterra-model}, leads to (quantum) fragmentation of the Hilbert space into exponentially many dynamically disconnected sectors~\cite{Moudgalya_2022}. In Section~\ref{ssec:chiral-Potts-fragments}, we will invoke a local basis transformation to map the sectors of the quantum Volterra model directly into a family of inhomogeneous quantum chiral Potts models. Surprisingly, and as far as we are aware, such a direct connection has not been reported before. In Section~\ref{ssec:integrability}, we will then review how the integrable structure of the quantum Volterra model reduces within each ``Potts sector''. This reduction of the integrable structure has been known in the context of a closely related lattice sine--Gordon model~\cite{Bazhanov_1990,Faddeev_1992,Bazhanov_1996,Bazhanov_2008,Faddeev_1996}.

\subsection{Fragments and quantum chiral Potts models}
\label{ssec:chiral-Potts-fragments}

To establish Hilbert space fragmentation, we resolve the localized constraints and the on-site symmetries by constructing a local basis that diagonalizes some maximal Abelian subset of $\{\tilde{w}_n\}_{n=1}^{2L}\cup\{U_1,U_2\}$. 

We start with $uv^{-1}\ket{\psi_j}=q^j\ket{\psi_j}$ and $uv\ket{\chi_j}=q^j\ket{\chi_j}$, where
\begin{align}
\begin{aligned}
    &\ket{\psi_j}=\frac{q^{1+j}\ket{0}+q^{1-j}\ket{1}+\ket{2}}{\sqrt{3}},
    \\
    &\ket{\chi_j}=\frac{q^{-j}\ket{0}+q^{1+j}\ket{1}+\ket{2}}{\sqrt{3}},
\end{aligned}
\end{align} 
and $\{\ket{0},\ket{1},\ket{2}\}$ is the ordered basis in which $u$ and $v$ take the matrix form in Eq.~\eqref{eq:u_v-matrices}. Consider a set of states $\{\ket{\psi_j}\otimes \ket{\chi_{s-j}}\}_{j\in\mathbb{Z}_3}$ for a fixed label $s\in\mathbb{Z}_3$ (the indices and labels are $0,1,2$ modulo $3$). Within such a set, the operators $u\otimes u$, $v^{-1}\otimes v$, and $\tilde{w}$ act as permutations. To diagonalize them, we then take Fourier transforms
\begin{align}
    \ket{\phi_{s,c}}=\sum_{j\in\{0,1,2\}}\frac{q^{jc}}{\sqrt{3}}\ket{\psi_j}\otimes \ket{\chi_{s-j}},
\end{align}
which obey
\begin{align}
    (u\otimes u)\ket{\phi_{s,c}}&=q^{1-c}\ket{\phi_{s,c}}, \label{eq:uu-eigenvalue} \\ 
    (v^{-1}\otimes v)\ket{\phi_{s,c}}&=q^{s+c-1}\ket{\phi_{s,c}}, \label{eq:vv-eigenvalue} \\
    \tilde{w}\ket{\phi_{s,c}}&=q^s \ket{\phi_{s,c}}.
    \label{eq:tilde-w-resolution}
\end{align}

Fragmentation now becomes explicit in the basis of states $\ket{\boldsymbol{s},\boldsymbol{c}}\equiv \bigotimes_{n=1}^{L}\ket{\phi_{s_n,c_n}}$, where $\boldsymbol{s}=(s_1,\ldots,s_{L})\in\mathbb{Z}_3^{L}$ is the sector label, determining the eigenvalues of the resolved constraints $\{\tilde{w}_{2n}\}$, and $\boldsymbol{c}=(c_1,\ldots,c_{L})\in\mathbb{Z}_3^L$ runs through configurations (or states) within the sector. In this basis, the local Hamiltonian densities~\eqref{eq:q-Volterra-model} act as
\begin{align}
    h_{2n-1,2n}\!\ket{\boldsymbol{s},\!\boldsymbol{c}}\!=\!-\!\frac{q^{s_n+1-c_n}+q^{c_n-1-s_n}}{\sqrt{3}}\!\ket{\boldsymbol{s},\!\boldsymbol{c}}
    \label{eq:h_even-action}
\end{align}
and
\begin{align}
    h_{2n,2n+1}\!\ket{\boldsymbol{s},\!\boldsymbol{c}}\!=&-\!\frac{q^{-s_n}}{\sqrt{3}}\!\ket{\boldsymbol{s},\!(\ldots,c_n\!+\!1,c_{n+1}\!-\!1,\ldots)}
    \notag\\
    &-\!\frac{q^{s_{n}}}{\sqrt{3}}\!\ket{\boldsymbol{s},\!(\ldots,c_n\!-\!1,c_{n+1}\!+\!1,\ldots)},
    \label{eq:h_odd-action}
\end{align}
for $n\in\{1,\ldots,L\}$. Note that the action of $H_{\rm V}$ preserves the sector label $\boldsymbol{s}$, and the Hilbert space therefore decomposes into $3^{L}$ dynamically disconnected components of equal size. Within each sector, we can compactly rewrite the Hamiltonian using two new $\mathbb{Z}_3$ clock operators
\begin{align}
    Z=
    \begin{pmatrix}
        q & 0 & 0 \\
        0 & 1 & 0 \\
        0 & 0 & q^{-1}
    \end{pmatrix},
    \qquad
    X=\begin{pmatrix}
        0 & 0 & 1 \\
        1 & 0 & 0 \\
        0 & 1 & 0
    \end{pmatrix},
    \label{eq:Z_X-matrices}
\end{align} 
which act only within the space of configurations $\boldsymbol{c}$, and which satisfy Weyl relation $ZX=q^{-1}XZ$. Using Eqs.~\eqref{eq:h_even-action} and~\eqref{eq:h_odd-action}, we find
\begin{align}
    H^{(\boldsymbol{s})}_{\rm P}
    =-\frac{1}{\sqrt{3}}\sum_{n=1}^{L}q^{s_n}(Z^{}_n  + X^{-1}_{n}X^{}_{n+1})  +  {\rm H.c.},
    \label{eq:chiral-Potts}
\end{align}
where $H^{(\boldsymbol{s})}_{\rm P}=\mathcal{P}^{}_{\boldsymbol{s}} H^{}_{\rm V}\mathcal{P}^\dagger_{\boldsymbol{s}}$, and $\mathcal{P}_{\boldsymbol{s}}=\sum_{\boldsymbol{c}}\ket{\boldsymbol{c}}\!\bra{\boldsymbol{s},\boldsymbol{c}}$ maps the sector with a fixed $\boldsymbol{s}$ into a reduced Hilbert space of dimension $3^L$. Equations~\eqref{eq:chiral-Potts} describe a family of inhomogeneous quantum $\mathbb{Z}_3$ chiral Potts models---one for each sector $\boldsymbol{s}$. We emphasize that the mapping of the Volterra model to the chiral Potts family presented here is based entirely on local basis transformations. To the best of our knowledge, no previous work has reported such a direct mapping. In particular, our correspondence does not invoke the models' integrable structures, which are known to be related (see, e.g., Ref.~\cite{Bazhanov_2008}).\footnote{Our correspondence is algebraically reminiscent of the construction recently used in Ref.~\cite{Li_2026}, where an $L$-site Baxter--Fendley model is rewritten in terms of $2L$ composite operators obeying Weyl algebra. The correspondence in Ref.~\cite{Li_2026} proceeds in the opposite direction and is formulated only at the operator level, without an explicit basis transformation or resolution of localized constraints.}

Finally, we consider the on-site symmetries~\eqref{eq:global-symmetries} within each sector. Using Eqs.~\eqref{eq:uu-eigenvalue} and~\eqref{eq:vv-eigenvalue} we find
\begin{align}
    \mathcal{P}^{}_{\boldsymbol{s}}U^{}_{1}\mathcal{P}^\dagger_{\boldsymbol{s}} = U_Z\equiv \prod_{n=1}^{L}Z_n,
    \quad \mathcal{P}^{}_{\boldsymbol{s}}U^{}_{2}\mathcal{P}^\dagger_{\boldsymbol{s}}=q^{S} U_Z^{-1},
    \label{eq:mapped-Fs}
\end{align}
where we have denoted $S=\sum_{n=1}^{L}s_n \mod 3$. Notably, $U_Z$ is diagonal in the configurational basis, and it does not depend on the label $\boldsymbol{s}$. Since $[U_Z,H_{\rm P}^{(\boldsymbol{s})}]=0$, each chiral Potts Hamiltonian~\eqref{eq:chiral-Potts} further decomposes into three blocks with a fixed $C=\sum_{n=1}^{L}c_n \mod 3$.

Under mild assumptions on the system size, we will now show that the inhomogeneous chiral Potts models~\eqref{eq:chiral-Potts} can be made homogeneous via a unitary transformation. For $\boldsymbol{a}=(a_1,\ldots,a_L)\in\mathbb{Z}_3^L$ and $\boldsymbol{b}=(b_1,\ldots,b_L)\in\mathbb{Z}_3^L$, we introduce the on-site operator~\footnote{Note that $U_Z=U_{\boldsymbol{1},\boldsymbol{0}}$, where $\boldsymbol{1}=(1,\ldots,1)$ and $\boldsymbol{0}=(0,\ldots,0)$ are $L$-dimensional vectors of ones and zeros, respectively.}
\begin{align}
    U_{\boldsymbol{a},\boldsymbol{b}}=\prod_{n=1}^{L}Z_n^{a_n}X_n^{b_n},
\end{align}
with the following adjoint actions on the terms of the Hamiltonian~\eqref{eq:chiral-Potts}: $U_{\boldsymbol{a},\boldsymbol{b}}^{} Z_n^{} U_{\boldsymbol{a},\boldsymbol{b}}^{-1}= q^{b_n}Z_n^{}$ and $U_{\boldsymbol{a},\boldsymbol{b}}^{} X_n^{-1}X_{n+1}^{} U_{\boldsymbol{a},\boldsymbol{b}}^{-1}= q^{a_n-a_{n+1}}X_n^{-1}X_{n+1}^{}$.

Applying such a transformation to Eq.~\eqref{eq:chiral-Potts}, the inhomogeneous phases $q^{s_n}$ can all be replaced by $q^s$, provided that
\begin{align}
    q^{b_n+s_n}\!=\!q^s\!=\!q^{a_n-a_{n+1}+s_n},\qquad a_{L+1}\!=\!a_1.
    \label{eq:transform-conditions}
\end{align}
Under such conditions we find $U_{\boldsymbol{a},\boldsymbol{b}}^{}H_{\rm P}^{(\boldsymbol{s})}U_{\boldsymbol{a},\boldsymbol{b}}^{-1}=H_{\rm P}^{(s)}$, where
\begin{align}
    H_{\rm P}^{(s)}=-\frac{1}{\sqrt{3}}\sum_{n=1}^{L}q^{s} (Z^{}_n  +  X^{-1}_{n}X^{}_{n+1})  +  {\rm H.c.}
    \label{eq:homogeneous-Potts}
\end{align}
is a translationally invariant quantum chiral Potts model, and $s\in\{0,1,2\}$. The conditions~\eqref{eq:transform-conditions}, under which such a mapping is valid, are satisfied by setting $b_n=s-s_n \mod 3$ and $a_n-a_{n+1}=b_n \mod 3$ for every $n$. Along with $a_{L+1}=a_1$ this leads to 
\begin{align}
    L s - \sum_{n=1}^L s_n\quad(\text{mod }3) = 0.
    \label{eq:s-condition}
\end{align}
Crucially, if $L$ is not divisible by $3$, we can always choose the phase $s\in\{0,1,2\}$ so as to satisfy the condition~\eqref{eq:s-condition}. Hence, for $3\nmid L$, any model in the family~\eqref{eq:chiral-Potts} is equivalent to some translationally invariant quantum chiral Potts model of the form given in Eq.~\eqref{eq:homogeneous-Potts}. It is reasonable to expect that limiting to system sizes which are not divisible by $3$ has no consequences for the bulk physics in the thermodynamic limit, e.g., for the model's transport properties.

\subsection{Integrability in full Hilbert space and within ``Potts'' sectors}
\label{ssec:integrability}

To place our mapping above in the context of existing results, 
we will briefly review the quantum Volterra model's integrability and how it reduces when the localized constraints are resolved. Aside from providing some historical context, the integrable structure will later help us establish ballistic energy transport in the quantum Volterra model.

Up to a term proportional to an identity, the Hamiltonian $H_{\rm V}$ in Eq.~\eqref{eq:q-Volterra-model} is equivalent to the logarithmic derivative $({\rm d}/{\rm d} \lambda) \log T(\lambda)\mid_{\lambda=0}$ of the transfer matrix
\begin{align}
    T(\lambda)={\rm Tr}_a\left[R_{1,a}(\lambda)\cdots R_{2L,a}(\lambda)\right],
    \label{eq:Volterra-transfer-matrix-1}   
\end{align}
where $a$ denotes the auxiliary copy of a three-dimensional local physical space, and $R(\lambda)$ is the $R$ matrix satisfying the Yang-Baxter equation
\begin{align}
R_{a_1,a_2}(\lambda-\mu)&R_{n,a_2}(\lambda)R_{n,a_1}(\mu)=
\notag\\
&=R_{n,a_1}(\mu)R_{n,a_2}(\lambda)R_{a_1,a_2}(\lambda-\mu).
\end{align}
The latter enforces the commutation of the transfer matrices~\eqref{eq:Volterra-transfer-matrix-1} for different values of the spectral parameter $\lambda$, i.e., $[T(\lambda),T(\mu)]=0$.  

We can express the $R$ matrix acting on neighboring sites $1$ and $2$ as
\begin{align}
    R_{1,2}(\lambda)=P_{1,2}\check{R}_{1,2}(\lambda),
    \label{eq:R-matrix}
\end{align}
where $P=\sum_{j,l\in\{0,1,2\}}\ket{j,l}\!\bra{l,j}$ is a permutation, and
\begin{align}
    \check{R}_{1,2}(\lambda)=e^{-i \tau(\lambda) \left[h_{1,2}-\mathbbm{1}/\sqrt{3}\right]}
    \label{eq:braid-R}
\end{align}
is a unitary braid form of the $R$ matrix. The latter is compactly expressed in terms of the Hamiltonian density~\eqref{eq:q-Volterra-model} and
\begin{align}
    \tau(\lambda)=\frac{i}{\sqrt{3}}\log\left[\frac{\coth(\frac{\lambda}{2})-i\sqrt{3}}{\coth(\frac{\lambda}{2})+i\sqrt{3}}\right].
    \label{eq:tau-func}
\end{align}
Since $\tau(\lambda)=\lambda+O(\lambda^3)$, the braid-form $R$ matrix~\eqref{eq:braid-R} can be used as a quantum gate in the integrable Trotterization of the unitary evolution $e^{-itH_{\rm V}}$~\cite{Faddeev_1994,Faddeev_1996,Zadnik_2017,Zadnik_2018}. Such an integrable Trotterization plays a prominent role as the lattice discretization of the sine--Gordon model~\cite{Faddeev_1992,Bazhanov_1996,Bazhanov_2008,Faddeev_1994,Faddeev_1996}. Its integrability structure is related to that of the XXZ-type model with cyclic spins, and one can leverage this to obtain the Bethe equations for the lattice sine--Gordon model, as well as to observe the connection to the chiral Potts model (e.g., see Refs.~\cite{Bazhanov_1990,Faddeev_1996,Bazhanov_2008}). 

To illustrate the connection between the integrability of the quantum Volterra model and that of the XXZ-type cyclic-spin model, we consider another transfer matrix. The crucial difference from the one in Eq.~\eqref{eq:Volterra-transfer-matrix-1} lies in the auxiliary space, which is now two-dimensional instead of three-dimensional. Labeling the two-dimensional auxiliary space by $\mathfrak{b}$, the new transfer matrix reads
\begin{align}
    \tilde{T}(\lambda)\!=\!{\rm Tr}_\mathfrak{b}\left[\operatorname{L}_{1,\mathfrak{b}}(\lambda)\cdots\operatorname{L}_{2L,\mathfrak{b}}(\lambda)\right].
    \label{eq:Volterra-transfer-matrix-2}
\end{align}
We have $[\tilde{T}(\lambda),T(\mu)]=0$ and $[\tilde{T}(\lambda),\tilde{T}(\mu)]=0$ as a courtesy of the Yang-Baxter equation
\begin{align}
    R_{n,a}(\lambda-\mu)&\operatorname{L}_{n,\mathfrak{b}}(\lambda)\operatorname{L}_{a,\mathfrak{b}}(\mu)
    =
    \notag\\
    &=\operatorname{L}_{a,\mathfrak{b}}(\lambda)\operatorname{L}_{n,\mathfrak{b}}(\lambda)R_{n,a}(\lambda-\mu),
\end{align}
satisfied by the $R$ matrix~\eqref{eq:R-matrix} and Lax operator
\begin{align}
    \operatorname{L}_{a,\mathfrak{b}}(\lambda)=&\,u^{}_a \frac{\mathbbm{1}+\sigma^z_\mathfrak{b}}{2}-e^\lambda v_a^{-1}\sigma_\mathfrak{b}^+ 
    \notag\\
    &+e^\lambda v^{}_a \sigma^-_\mathfrak{b}+u_a^{-1} \frac{\mathbbm{1}-\sigma^z_\mathfrak{b}}{2},
\end{align}
where $\sigma^{\pm,z}_\mathfrak{b}$ are the Pauli matrices.

To map the transfer matrix~\eqref{eq:Volterra-transfer-matrix-2} into the one of the cyclic-spin XXZ model, we resolve the local non-Abelian constraints, $\tilde{w}_{2n}\to q^{s_n}$. Using the auxiliary-space gauge matrices $G^{(n)}={\rm diag}(z^{}_n,z_n^{-1})$, with $z_{n+1}=z_n q^{-s_n/2}$, we then obtain
\begin{align}
    G_\mathfrak{b}^{(n)}\operatorname{L}^{}_{2n-1,\mathfrak{b}}(\lambda)\operatorname{L}^{}_{2n,\mathfrak{b}}(\lambda)\left[G_\mathfrak{b}^{(n+1)}\right]^{-1}=
    \notag\\
    =e^\lambda (q-q^{-1}) \mathbbm{L}_{n,\mathfrak{b}}(\lambda),
\end{align}
where
\begin{align}
    \mathbbm{L}_{n,\mathfrak{b}}(\lambda)=
    \begin{pmatrix}
        \frac{e^{-\lambda} q^{S_n^z}-e^{\lambda}q^{-S_n^z}}{q-q^{-1}} & S_n^- \\
        S_n^+ & \frac{e^{-\lambda}q^{-S_n^z}-e^{\lambda} q^{S_n^z}}{q-q^{-1}}
    \end{pmatrix}
\end{align}
is the Lax operator of the cyclic-spin XXZ model. In particular, the cyclic-spin operators which act on the physical degrees of freedom are
\begin{align}
\begin{aligned}
	q^{S_n^z}&= q^{-s_n/2}u_{2n-1}u_{2n},\\
	S_n^+&= z_n z_{n+1}\,\frac{u^{-1}_{2n-1}v^{}_{2n}+v^{}_{2n-1}u^{}_{2n}}{q-q^{-1}},\\
	S_n^-&=-\frac{1}{z_n z_{n+1}}\,\frac{u^{}_{2n-1}v^{-1}_{2n}+v^{-1}_{2n-1}u^{-1}_{2n}}{q-q^{-1}},
	\label{eq:spin-ops}
\end{aligned}
\end{align}
and they satisfy the $\mathcal{U}_q(sl_2)$ algebra
\begin{align}
	[S_n^+,S_n^-]\!=\!\frac{q^{2S_n^z}\!-\!q^{-2S_n^z}}{q\!-\!q^{-1}},
    \quad
    q^{S_n^z}S_n^{\pm}\!=\!q^{\pm1}S_n^{\pm }q^{S_n^z},
\end{align}
for each $n\in\{1,\ldots,L\}$ separately (for different $n$ they commute).
In this way, the transfer matrix~\eqref{eq:Volterra-transfer-matrix-2} becomes equivalent to a twisted XXZ-type transfer matrix
\begin{align}
\tilde{T}(\lambda)
\!=\!
\left[e^\lambda (q\!-\!q^{-1})\!\right]^L \!{\rm Tr}_\mathfrak{b}\!\!\left[\mathbbm{L}_{1,\mathfrak{b}}(\lambda)\cdots \mathbbm{L}_{L,\mathfrak{b}}(\lambda)\mathbbm{G}_{\mathfrak{b}}^{(\boldsymbol{s})}\right]\!,
\end{align}
in which the boundary twist
\begin{align}
    \mathbbm{G}^{(\boldsymbol{s})}&=G^{(L+1)}[G^{(1)}]^{-1}
    \notag\\
    &={\rm diag}\left(q^{-\!\frac{1}{2}\!\sum_{m=1}^L \!s_m},q^{\frac{1}{2}\!\sum_{m=1}^L \! s_m}\right)
\end{align}
is completely determined by the sector label $\boldsymbol{s}$. In the context of lattice sine--Gordon model, the above mapping has been used for $\boldsymbol{s}=\boldsymbol{0}$~\cite{Bazhanov_1990,Faddeev_1996}.

\section{Nonintegrable classical limit}
\label{sec:not-a-gauge-theory}

The fact that the dynamics within all fragments of the quantum Volterra model admit a unified description might suggest an underlying lattice gauge structure. To examine this possibility, we turn to the classical Hamiltonian setting, where the nature of the constraints can be established by analyzing their Poisson-bracket algebra. 

We consider a nonintegrable classical limit of the quantum Volterra model, obtained by setting 
\begin{align}
    u_n=e^{i\varphi_n},\qquad v_n=e^{i\pi_n},
\end{align}
where $\varphi_n$ and $\pi_n$ denote classical field and momentum variables, respectively. Operator $u_n$ can then be interpreted as a boost and $v_n$ as a shift. In the classical limit, the Weyl algebra~\eqref{eq:Weyl-relation} reduces to the standard Poisson bracket
\begin{align}
    \{\varphi_n,\pi_m\} = \delta_{n,m},
\end{align}
while the Hamiltonian~\eqref{eq:q-Volterra-model} becomes
\begin{align}
    H_{\rm cl}\!=\!\sum_{n=1}^{2L}\cos(\varphi_n+\pi_n+\varphi_{n+1}-\pi_{n+1}),
\label{eq:classical-model}
\end{align}
up to a prefactor which will not be relevant for our discussion. Note that the classical dynamics generated by the Hamiltonian~\eqref{eq:classical-model} are not integrable. Indeed, since $u_n$ and $v_n$ can take any value on the unit circle, our classical limit can be viewed as the $N\to\infty$ limit ($q^N=1$) of the quantum Hamiltonian~\eqref{eq:q-Volterra-model}, taken without modifying its form. It is known, however, that additional terms must be added to $H_V$ in order to preserve its integrability for higher-order roots-of-unity~\cite{Volkov_1992}.

The Hamiltonian equations of motion read
\begin{align}
\frac{{\rm d}\varphi_n}{{\rm d}t} \! =\!\sin f_{n-1} \! - \! \sin f_n,
\quad
\frac{{\rm d}\pi_n}{{\rm d}t}\!=\!\sin f_{n-1}  \!+ \!\sin f_n,
\label{eq:classical-EoM}
\end{align}
where $n\in\{1,\ldots,2L\}$,
\begin{align}
    f_n:=\varphi_n+\pi_n+\varphi_{n+1}-\pi_{n+1},
    \label{eq:f-function}
\end{align}
and periodic boundary conditions are imposed. Combined, these equations yield a closed system of equations,
\begin{align}
    \frac{{\rm d}f_n}{{\rm d}t}
    = 2\sin f_{n-1} - 2\sin f_{n+1},
    \label{eq:Burgers-lattice}
\end{align}
determining the evolution of the local energy densities $h_{n,n+1}=\cos f_n$ [cf. Eq.~\eqref{eq:classical-model}]. We note that, in the continuum limit $a\to 0$, with $x=an$ and
$\tau=at$ held fixed, this leads to the inviscid Burgers' equation for the energy density,
\begin{align}
\frac{\partial }{\partial\tau}h(x,t)
+ 2\frac{\partial }{\partial x}h^2(x,t) = 0,
\label{eq:Burgers}
\end{align}
which takes the standard form upon rescaling the space coordinate as $x\to 4x$.

\subsection{The role of localized constraints}

To determine whether we are dealing with a lattice gauge structure, we now analyze the constraints. In particular, the dual variables $\{\tilde{w}_n\}$ generate a set of localized constraints imposed on the fields and momenta. The dual variables can be expressed as 
\begin{align}
    \tilde{w}_n = e^{i \sigma_n}, 
\end{align}
where
\begin{align}
    \sigma_n := \varphi_n - \pi_n + \varphi_{n+1} + \pi_{n+1}
    \label{eq:classical-constraints}
\end{align}
are conserved: by construction one has $\{f_n,\sigma_m\}=0$ and thus $\{H_{\rm cl},\sigma_m\}=0$. As a result, a classical trajectory $(\boldsymbol{\varphi}(t),\boldsymbol{\pi}(t))$ is constrained to a submanifold of the full phase space, which replaces the notion of a fragment in the Hilbert space of the quantum Volterra model. 

We now argue that, while the local constraints~\eqref{eq:classical-constraints} are extensive in number, they do not generate a lattice gauge
symmetry. Indeed, according to Ref.~\cite{Henneaux_1992}, proper gauge transformations are generated by the so-called \emph{first-class constraints}, whose Poisson brackets with all constraints vanish on the \emph{constraint surface} defined by setting all constraints to zero. Put differently, the transformations generated by first-class constraints preserve the constraint surface. A lattice gauge theory requires extensively many first-class constraints. To determine their number in our case, we consider the matrix of Poisson brackets $C_{m,n}=\{\sigma_n,\sigma_m\}$~\cite{Canepa_2021}, which reads
\begin{equation}
C_{m,n} = 2\delta_{n,m+1}
       - 2\delta_{n,m-1},
\label{eq:Cmatrix}
\end{equation}
with indices taken modulo $2L$. Since ${\rm dim}({\rm ker}\,C)=2$, there are only two independent first-class combinations of the constraints, which precludes the interpretation of $\{\sigma_n\}$ as generators of a lattice gauge symmetry.

\section{Localized constraints and transport}
\label{sec:energy-transport}

Local conserved quantities constrain the dynamics and are therefore expected to affect charge transport. This connection can be made quantitative: transport coefficients such as the Drude weight, which characterizes ballistic transport, and the diffusion constant can be bounded or expressed using the overlaps between conserved quantities and the charge current~\cite{Zotos_1997,Prosen_2014,Doyon_2017,Medenjak_2017}. The natural question in the present context is what role uniform fragmentation and the associated localized constraints play in the transport properties.

\subsection{Energy transport in quantum Volterra model}

We first consider energy transport in the quantum Volterra model, where the hierarchy of local conserved quantities generated by the transfer matrix~\eqref{eq:Volterra-transfer-matrix-1} contains the energy current itself. The conserved quantities are given by the logarithmic derivatives 
$Q_n=({\rm d}/{\rm d} \lambda)^n \log T(\lambda)\mid_{\lambda=0}$. The first one is the Hamiltonian, $Q_1=H_{\rm V}$, while the second one is the energy current. To see this, we write the second logarithmic derivative as
\begin{align}
    Q_2=\sum_{n=1}^{2L}\check{R}''_{n,n+1}-(\check{R}'_{n,n+1})^2
    +[\check{R}'_{n-1,n},\check{R}'_{n,n+1}],
\end{align}
where $(\bullet)':=({\rm d}/{\rm d}\lambda)(\bullet)_{\lambda=0}$. For the $R$ matrix~\eqref{eq:braid-R} we have $\check{R}''=(\check{R}')^2$, and we obtain
\begin{align}
    Q_2=-\sum_{n=1}^{2L}[h_{n-1,n},h_{n,n+1}],
    \label{eq:conserved-current}
\end{align}
which is the total energy current. Indeed, from the continuity equation
\begin{align}
    \frac{\partial}{\partial t}h_{n-1,n}(t)+j_n(t)-j_{n-1}(t)=0
\end{align}
we identify the energy current density as $j_n=i[h_{n-1,n},h_{n,n+1}]$. Its explicit form reads
\begin{equation}
j_n
\!=\!
\frac{1}{\sqrt{3}}
u_{n-1}^{}v_{n-1}^{}
\big(
v_n^{}u_{n+1}^{-1}v_{n+1}^{}
\!-\!
u_n^{-1}u_{n+1}^{}v_{n+1}^{-1}
\big)
+\text{H.c.}
\label{eq:current-explicit}
\end{equation}
Eq.~\eqref{eq:conserved-current} thus implies that the total energy current is conserved. The same feature has been observed in the classical limit of the model~\cite{Mazzuca_2024}.

Since the energy current is a conserved quantity, the energy transport in the quantum Volterra model is ballistic. We will focus on the energy transport at infinite temperature. It is characterized by a finite Drude weight
\begin{align}
    D_V=\lim_{t\to\infty}\sum_{n\in\mathbb{Z}} \left(\frac{n}{t}\right)^2 S_V(n,t),
    \label{eq:Drude-V}
\end{align}
where 
\begin{equation}
S(n,t):=\langle h_{n-1,n}(t),h_{-1,0}(0)\rangle_c
\label{def:SF}
\end{equation}
is the energy dynamical structure factor, the label $V$ in Eq.~\eqref{eq:Drude-V} referring to the quantum Volterra model. Here, $\langle A,B\rangle_c=\langle A B\rangle-\langle A\rangle \langle B\rangle$ is the connected correlation function in the infinite-temperature state $\langle\bullet\rangle$, and we consider an infinite system, $n\in\mathbb{Z}$. Using the time-translation invariance of the  state, we can alternatively express the Drude weight in terms of local current autocorrelation functions
\begin{align}
    D_V=\lim_{t\to\infty}\sum_{n\in\mathbb{Z}}\langle j_{n}(t),j_{0}(0)\rangle_c.    
\end{align}
Since the current is conserved, this yields $D_V=\langle j_0,j_0\rangle_c=4/3$.

\begin{figure}[t!]
\centering
\includegraphics[width=1\linewidth]{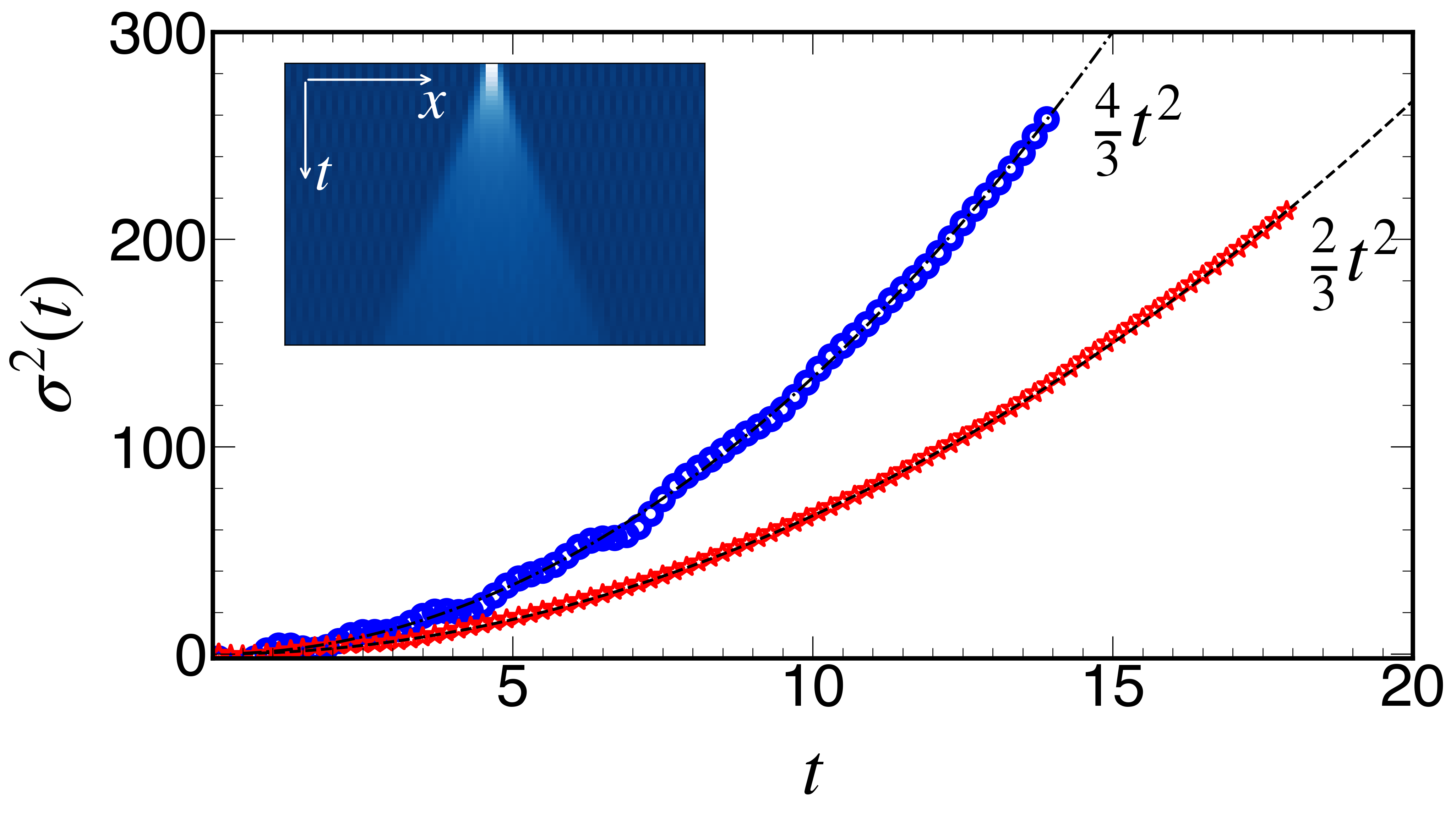}
\caption{\textbf{Ballistic energy transport in the quantum Volterra model.} The growth of the dynamical structure factor's variance $\sigma^2(t)=\sum_n n^2 S(n,t)$ in the quantum Volterra model (blue circles) and in the homogeneous chiral Potts model, Eq.~\eqref{eq:homogeneous-Potts}, with $s=1$ (red stars). Data was obtained using a TEBD method with maximal bond dimension $128$ and maximal allowed ratio between the largest and smallest Schmidt coefficient $10^6$. The time step in the first-order Trotter-Suzuki scheme was $\Delta t=0.02$, and system sizes were $L=161$ and $L=109$, respectively. Within linear response theory, the structure factor at time $t$ is related to the energy profile through $S(n,t) \approx (\langle h_{n-2,n-1}(t)\rangle-\langle h_{n-1,n}(t)\rangle)/(2\Delta\beta)$~\cite{Ljubotina_2019}, where the system has initially been prepared in a bipartite state with $\beta_{\rm left}=\Delta\beta$ and $\beta_{\rm right}=-\Delta\beta$. We have used $\Delta\beta=5\times 10^{-3}$.}
\label{fig:Drude}
\end{figure}

Consider now a homogeneous chiral Potts model~\eqref{eq:homogeneous-Potts}. For $3\nmid L$, which is a benign assumption in the thermodynamic limit, such a model effectively describes each fragment in the Hilbert space of $H_V$. Note that we obtained the local Hamiltonian density of the Potts model by projecting the sum of two neighbouring Hamiltonian densities of $H_V$, i.e., $h_{2n-1,2n}+h_{2n,2n+1}$ [see Eqs.~\eqref{eq:h_even-action} and~\eqref{eq:h_odd-action}]. From there, it follows that
\begin{align}
    D_P=\lim_{t\to\infty}\sum_{n\in\mathbb{Z}} \left(\frac{n}{t}\right)^2 S_P(n,t)
    =\frac{1}{2}D_V=\frac{2}{3},
    \label{eq:Drude-weights-relation}
\end{align}
which is indeed confirmed by tensor-network numerics, as reported in Fig.~\ref{fig:Drude}. This suggests that all fragments of the Hilbert space exhibit the same energy-transport properties and that the localized constraints associated with uniform fragmentation do not affect charge transport. This can be understood in the context of the model's integrability. In integrable systems, Drude weights are determined by the thermodynamic quasiparticle distributions, their effective velocities, and dressed charges~\cite{Doyon_2017}. These depend on the macrostate and can be computed using the thermodynamic Bethe ansatz within each fragment. In our case, each fragment is described by a (homogeneous) quantum Potts chain in Eq.~\eqref{eq:homogeneous-Potts}. The phases $q^s$ that distinguish the Potts models within different fragments can be transferred into a boundary twist using a local unitary transformation~\cite{Martins_2026}.\footnote{This transformation has a similar role as the one applied to the transfer matrix in Section~\ref{ssec:integrability}.} The twist introduces an additional term into the logarithmic form of the Bethe equations, which becomes irrelevant in the thermodynamic limit. Consequently, the bulk quasiparticle description is independent of the fragment, and so are the corresponding Drude weights evaluated in the same thermodynamic ensemble. In particular, distinct fragments do not specify distinct thermodynamic macrostates.

We next demonstrate that this feature, i.e., the independence of transport from the values of the localized constraints, is not specific to integrable quantum dynamics but extends to nonintegrable classical systems as well.

\begin{figure*}[t]
\centering
\includegraphics[width=0.8\linewidth]{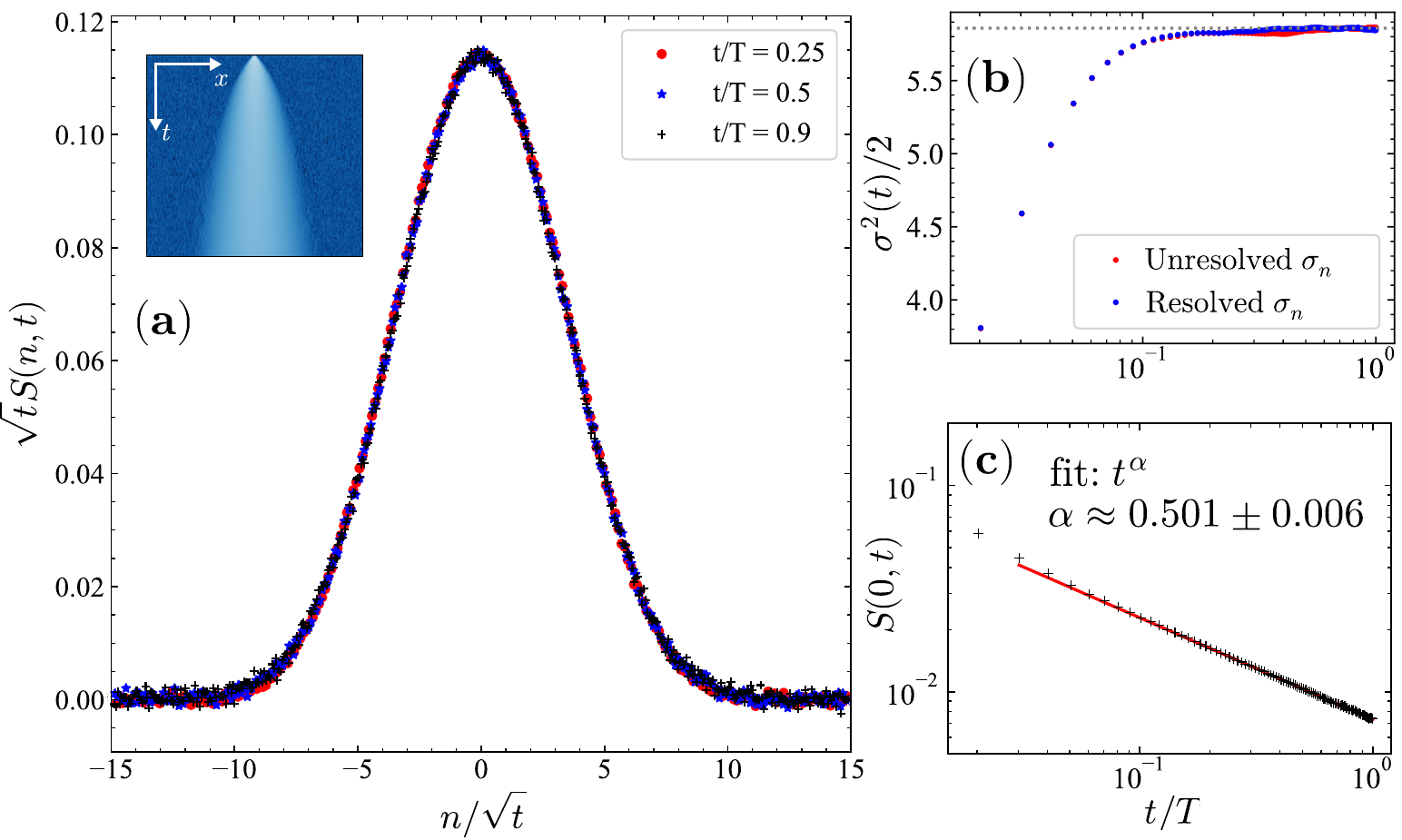}
\caption{\textbf{Diffusive energy transport in the classical limit.} Panel (a) shows the rescaled dynamical structure factor, Eq.~\eqref{def:SF}, as a function of $n/\sqrt{t}$, for different times $t$, exhibits a perfect collapse to a Gaussian form. The inset shows the evolution of $S(n,t)$ as a function of time (increasing from top to bottom) and space (left to right). Panel (b) shows the convergence of the variance of $S(n,t)$ to the diffusion constant (dotted horizontal line) as a function of $t/T$, where $T$ is the maximum simulation time. The blue dashed line is the result on a submanifold specified by fixed localized constraints $\{\sigma_n\}$, and the red solid line is the result obtained without resolving the constraints. Both diffusion constants are in perfect agreement. Panel (c) shows the decay of the peak of the structure factor, $S(0,t)$ (black points), and the resulting ``least squares fit'' (red solid line) showing $t^{\alpha}$ power law behaviour with $\alpha \simeq -0.501 \pm 0.006 $, signaling energy diffusion. The dynamics has been simulated in a system of $L = 5000$ sites, for a maximal time  $T = 500$. Averaging was performed over $10^5$ initial conditions sampled from the full phase space (without fixing the constraints) and over $10^5$ initial conditions restricted to a submanifold.}
\label{fig:Diffusion_classical}
\end{figure*}

\subsection{Energy diffusion in the classical model}

To distinguish the effects of the localized constraints from those of the local integrals of motion, we now turn to energy transport in a nonintegrable setting---the classical limit of the quantum Volterra model introduced in Section~\ref{sec:not-a-gauge-theory}.

We numerically compute the infinite-$T$ energy dynamical structure factor~\eqref{def:SF} for the classical dynamics encoded in Eq.~\eqref{eq:classical-EoM}. The structure factor and the scaling of its variance are reported in Fig.~\hyperref[fig:Diffusion_classical]{2(a)}. On large space-time scales, the structure factor assumes a Gaussian scaling form
\begin{align}
S(n,t) \simeq \frac1{\sqrt{4\pi{\cal D}t}} \exp\left(-\frac{n^2}{4{\cal D}t}\right),
\label{eq:Gaussian-scaling}
\end{align}
where $\mathcal{D}$ is the diffusion constant. We observe energy diffusion irrespective of whether localized constraints are resolved (infinite-$T$ ensemble on a submanifold with fixed $\{\sigma_n\}$) or not (unconstrained infinite-$T$ ensemble). In both cases, the structure factor is described by the same scaling function, Eq.~\eqref{eq:Gaussian-scaling}, with the same diffusion constant $\mathcal{D}$ [see Fig.~\hyperref[fig:Diffusion_classical]{2(b)}]. These results again indicate that the localized constraints do not affect the transport.

\section{Local ergodicity in uniformly-fragmented cellular automata}
\label{sec:thermalization}

Testing ergodicity requires determining how much of the phase space is accessible to single trajectories, which is a hard task in continuous-time dynamical systems. For this reason we will now focus on discrete-time classical dynamical systems with a finite configuration space. In Subsection~\ref{sec:automaton} we derive two cellular automata with discrete degrees of freedom, based on the Hamiltonian equations~\eqref{eq:classical-EoM}. In Subsection~\ref{sec:local-ergodicity} we then introduce the concept of local ergodicity~\cite{Orlov_2026}, which underlies the thermalization of local observables in discrete classical dynamical systems on a finite configuration space, and investigate how it breaks down due to the presence of localized constraints.

\subsection{Cellular automaton with localized constraints}
\label{sec:automaton}

To obtain a discrete-time dynamics with localized constraints, we start from a symplectic Euler integrator applied to the Hamiltonian dynamics in Eq.~\eqref{eq:classical-EoM}. Taking $\rho_n=\varphi_n+\pi_n$ and $\sigma_n$, Eq.~\eqref{eq:classical-constraints}, as variables, the localized constraints appear explicitly as parameters in the dynamics. The new variables satisfy 
\begin{align}
\begin{aligned}
    &\{\rho_n,\rho_m\}=0, \quad \{\rho_n,\sigma_m\}=-2\delta_{n,m}, 
    \\
    &\{\sigma_n,\sigma_m\}=2(\delta_{n,m+1}-\delta_{n,m-1}),
    \label{eq:Poisson_brackets}
\end{aligned}
\end{align}
and the Hamiltonian~\eqref{eq:classical-model} reads $H_{\rm cl}=\sum_{n=1}^{2L}\cos f(\rho_{n-1},\sigma_n,\rho_{n+1})$, where
\begin{align}
f(\rho_{n-1},\sigma_n,\rho_{n-1}):=\rho_{n-1}+\sigma_n-\rho_{n+1},
\label{def:Flin}
\end{align}
is simply an explicit redefinition of the variable introduced in Eq.~\eqref{eq:f-function}. The original variables can be recovered as $\varphi_n=(\rho_n-\rho_{n+1}+\sigma_n)/2$ and $\pi_n=(\rho_n+\rho_{n+1}-\sigma_n)/2$.

Representing the system as a horizontal staircase
\begin{center}
    \begin{tikzpicture}[scale=0.75]

\tikzset{
  caption1/.style={align=center, font=\fontsize{10pt}{10pt}},
  caption2/.style={align=center, font=\LARGE},
  caption3/.style={align=center, font=\fontsize{13pt}{10pt}}
}

\tikzset{
  pattern1/.style={fill=black!10},
  pattern2/.style={fill=black!0},
  pattern3/.style={red!50!white, line width=1.5pt},
  pattern4/.style={blue!50!white, line width=1pt}
}

\node[caption1] at (-1.5,1.5) {$\ldots$};
\node[caption1] at (7.5,1.5) {$\ldots$};

\draw[pattern1]
(0,1) -- (1,0) -- (2,1) -- (1,2) -- cycle;
\node[caption1] at (1,1) {$n-2$};

\draw[pattern1]
(2,1) -- (3,0) -- (4,1) -- (3,2) -- cycle;
\node[caption1] at (3,1) {$n$};

\draw[pattern1]
(4,1) -- (5,0) -- (6,1) -- (5,2) -- cycle;
\node[caption1] at (5,1) {$n+2$};

\draw[pattern1]
(-1,2) -- (0,1) -- (1,2) -- (0,3) -- cycle;
\node[caption1] at (0,2) {$n-3$};

\draw[pattern1]
(1,2) -- (2,1) -- (3,2) -- (2,3) -- cycle;
\node[caption1] at (2,2) {$n-1$};

\draw[pattern1]
(3,2) -- (4,1) -- (5,2) -- (4,3) -- cycle;
\node[caption1] at (4,2) {$n+1$};

\draw[pattern1]
(5,2) -- (6,1) -- (7,2) -- (6,3) -- cycle;
\node[caption1] at (6,2) {$n+3$};

\end{tikzpicture}
\end{center}
and applying the Euler symplectic integrator with a time step $\tau$ now yields a two-step discrete-time evolution with a local update rule
\begin{align}
\begin{aligned}
    \rho_{u}=\rho_{d}\!+\!2\tau \sin f(\rho_{\ell},\sigma_{d},\rho_{r}),\quad \sigma_u=\sigma_d,
    \label{eq:symplectic-euler}
\end{aligned}
\end{align}
applied as shown in Fig.~\hyperref[fig:saw]{3(a)}. Such discrete-time dynamics preserves the Poisson brackets~\eqref{eq:Poisson_brackets}. In order to obtain a finite, discrete configuration space, we furthermore restrict the allowed values taken by the variables $\rho_n$ and $\sigma_n$. Setting $\tau=2\pi/(3\sqrt{3})$ and noting that $2\tau\sin z=z \mod 2\pi$ for $z\in (2\pi/3)\mathbb{Z}_3$, the automaton can be reformulated in terms of discrete variables $\rho_n,\sigma_n\in\mathbb{Z}_3$ ($n \in\{1,\ldots,2L\}$), with the local update rule simplified into
\begin{align}
    \rho_{u}=\rho_{d}+f(\rho_{\ell},\sigma_{d},\rho_{r}) \mod 3,
    \quad 
    \sigma_u=\sigma_d.
    \label{eq:cellular_automaton}
\end{align}
The linearity makes such dynamics  unsuitable for identifying the ergodic properties that are lost due to localized constraints. Because of that, we introduce additional nonlinearities by making the local update depend also on the neighbouring plaquettes, as shown in Fig.~\hyperref[fig:saw]{3(b)}:
\begin{widetext}
\begin{align}
    \rho_u=
    \rho_d+f(\rho_\ell,\sigma_d,\rho_r)
    \big[1+f(\rho_{\ell\ell},\sigma_{\ell d},\rho_{\ell})+f(\rho_r,\sigma_{rd},\rho_{rr})\big] 
    \mod 3,
    \quad 
    \sigma_u=\sigma_d.
    \label{eq:nonlin-automaton}
\end{align}
\end{widetext}
The additional nonlinear terms preserve the symplectic property of the automaton and render the dynamics ergodic once all of the parameters $\{\sigma_n\}$ are fixed. This ensures that the localized constraints are the only source of nonergodicity.

\begin{figure}[t]
\centering
\begin{tikzpicture}[scale=0.75]

\tikzset{
  caption1/.style={align=center, font=\fontsize{10pt}{10pt}},
  caption3/.style={align=center, font=\fontsize{13pt}{10pt}}
}

\tikzset{
  pattern1/.style={fill=black!10},
  pattern2/.style={fill=black!0},
  pattern3/.style={red!40!white, line width=3pt},
  pattern4/.style={blue!40!white, line width=1.5pt}
}


\begin{scope}[shift={(0,6)}]

\node[font=\large] at (-2,0.3)
{\bf{(a)}};

\node[caption1] at (-1.5,1.5) {$\ldots$};
\node[caption1] at (-1.5,3.5) {$\ldots$};
\node[caption1] at (7.5,1.5) {$\ldots$};
\node[caption1] at (7.5,3.5) {$\ldots$};

\draw[pattern1]
(0,1) -- (1,0) -- (2,1) -- (1,2) -- cycle;

\draw[pattern1]
(2,1) -- (3,0) -- (4,1) -- (3,2) -- cycle;
\node[caption1] at (3,1) {$d$};

\draw[pattern1]
(4,1) -- (5,0) -- (6,1) -- (5,2) -- cycle;

\draw[pattern1]
(-1,2) -- (0,1) -- (1,2) -- (0,3) -- cycle;

\draw[pattern1]
(1,2) -- (2,1) -- (3,2) -- (2,3) -- cycle;
\node[caption1] at (2,2) {$\ell$};

\draw[pattern1]
(3,2) -- (4,1) -- (5,2) -- (4,3) -- cycle;
\node[caption1] at (4,2) {$r$};

\draw[pattern1]
(5,2) -- (6,1) -- (7,2) -- (6,3) -- cycle;

\draw[pattern2]
(0,3) -- (1,2) -- (2,3) -- (1,4) -- cycle;

\draw[pattern2]
(2,3) -- (3,2) -- (4,3) -- (3,4) -- cycle;
\node[caption1] at (3,3) {$u$};

\draw[pattern2]
(4,3) -- (5,2) -- (6,3) -- (5,4) -- cycle;

\draw[pattern2]
(-1,4) -- (0,3) -- (1,4) -- (0,5) -- cycle;

\draw[pattern2]
(1,4) -- (2,3) -- (3,4) -- (2,5) -- cycle;

\draw[pattern2]
(3,4) -- (4,3) -- (5,4) -- (4,5) -- cycle;

\draw[pattern2]
(5,4) -- (6,3) -- (7,4) -- (6,5) -- cycle;

\draw[pattern3]
(0.8,2) -- (3,-0.2) -- (5.2,2) -- (3,4.2) -- cycle;

\draw[pattern4]
(1.9,3) -- (4,0.9) -- (6.1,3) -- (4,5.1) -- cycle;

\end{scope}

\node[font=\large] at (-2,0.3)
{\bf{(b)}};

\node[caption1] at (-1.5,1.5) {$\ldots$};
\node[caption1] at (-1.5,3.5) {$\ldots$};
\node[caption1] at (7.5,1.5) {$\ldots$};
\node[caption1] at (7.5,3.5) {$\ldots$};

\draw[pattern1]
(0,1) -- (1,0) -- (2,1) -- (1,2) -- cycle;
\node[caption1] at (1,1) {$\ell d$};

\draw[pattern1]
(2,1) -- (3,0) -- (4,1) -- (3,2) -- cycle;
\node[caption1] at (3,1) {$d$};

\draw[pattern1]
(4,1) -- (5,0) -- (6,1) -- (5,2) -- cycle;
\node[caption1] at (5,1) {$rd$};

\draw[pattern1]
(-1,2) -- (0,1) -- (1,2) -- (0,3) -- cycle;
\node[caption1] at (0,2) {$\ell\ell$};

\draw[pattern1]
(1,2) -- (2,1) -- (3,2) -- (2,3) -- cycle;
\node[caption1] at (2,2) {$\ell$};

\draw[pattern1]
(3,2) -- (4,1) -- (5,2) -- (4,3) -- cycle;
\node[caption1] at (4,2) {$r$};

\draw[pattern1]
(5,2) -- (6,1) -- (7,2) -- (6,3) -- cycle;
\node[caption1] at (6,2) {$rr$};

\draw[pattern2]
(0,3) -- (1,2) -- (2,3) -- (1,4) -- cycle;

\draw[pattern2]
(2,3) -- (3,2) -- (4,3) -- (3,4) -- cycle;
\node[caption1] at (3,3) {$u$};

\draw[pattern2]
(4,3) -- (5,2) -- (6,3) -- (5,4) -- cycle;

\draw[pattern2]
(-1,4) -- (0,3) -- (1,4) -- (0,5) -- cycle;

\draw[pattern2]
(1,4) -- (2,3) -- (3,4) -- (2,5) -- cycle;

\draw[pattern2]
(3,4) -- (4,3) -- (5,4) -- (4,5) -- cycle;

\draw[pattern2]
(5,4) -- (6,3) -- (7,4) -- (6,5) -- cycle;

\draw[pattern3]
(-1.2,2) -- (1,-0.2) -- (2,0.8) -- (3,-0.2) -- (4,0.8) -- (5,-0.2) -- (7.2,2) -- (6,3.2) -- (5,2.2) -- (3,4.2) -- (1,2.2) -- (0,3.2) -- cycle;

\draw[pattern4]
(-0.1,3) -- (2,0.9) -- (3,1.9) -- (4,0.9) -- (5,1.9) -- (6,0.9) -- (8.1,3) -- (7,4.1) -- (6,3.1) -- (4,5.1) -- (2,3.1) -- (1,4.1) -- cycle;

\end{tikzpicture}

\caption{\textbf{Cellular automata.} Panel (a) depicts the update of a horizontal staircase (light gray) in the upwards direction, using a local update rule from either Eq.~\eqref{eq:symplectic-euler} or Eq.~\eqref{eq:cellular_automaton}. The update rule is first simultaneously applied to all elementary plaquettes at the bottom (one of them denoted in red) and then at the top (one of them denoted in blue). Panel (b) shows a similar two-step protocol with a nonlinear update rule, Eq.~\eqref{eq:nonlin-automaton}, applied to an extended plaquette.}
\label{fig:saw}
\end{figure}
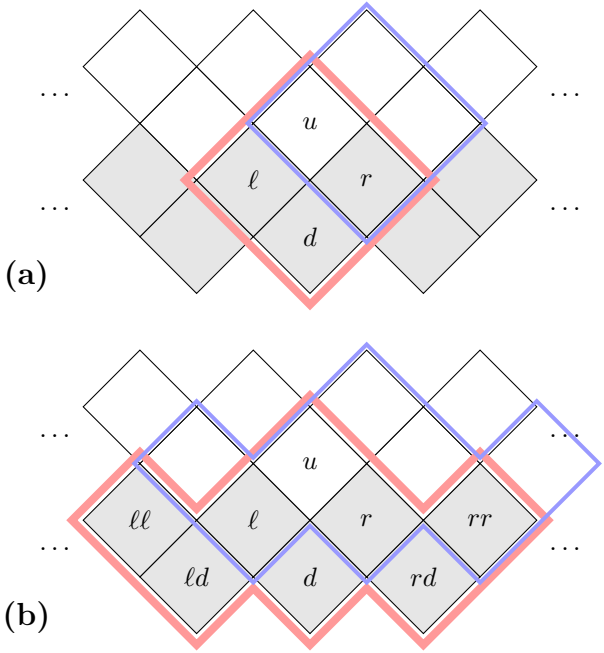

\subsection{Breakdown of local ergodicity due to constraints}
\label{sec:local-ergodicity}

The cellular automata considered herein are invertible deterministic maps $U:\mathcal{C}\to\mathcal{C}$ on a finite configuration space $\mathcal{C}=\mathbb{Z}_3^{(4L)}$, mapping between configurations of the form $\boldsymbol{x}=(\rho_1,\sigma_1,\rho_2,\sigma_2,\ldots,\rho_{2L},\sigma_{2L})$. While they could as well be thought of as maps between configurations $(\varphi_1,\pi_1,\varphi_2,\pi_2,\ldots \varphi_{2L},\pi_{2L})$, the representation in which the constraints $\{\sigma_n\}$ are explicit will serve our purpose better. 

Since $U^t$ is invertible and deterministic for any integer time $t$, it acts as a permutation and is thus decomposable into cycles (orbits), which generically precludes global ergodicity. However, as discussed in Ref.~\cite{Orlov_2026}, thermalization of local observables requires weaker versions of ergodicity and mixing. The former ensures that the orbit-averaged expectation values of \emph{local} observables match the microcanonical predictions, while the latter guarantees that the expectation values of local observables actually approach the corresponding predictions. Here, we focus only on the weak (or local) ergodicity. After briefly reviewing it, we will demonstrate that already this property fails to hold in the presence of localized constraints. 

We represent an observable $A$ as a function $\hat{A}:\mathcal{C}\to \mathbb{C}$, given by $\hat{A}=\sum_{\boldsymbol{x}} A(\boldsymbol{x}) \hat{I}_{\boldsymbol{x}}$, where $\hat{I}_{\boldsymbol{x}}(\boldsymbol{y})=\delta_{\boldsymbol{x},\boldsymbol{y}}$ denotes the indicator function. Similarly, a probability distribution on the configuration space can be represented as $\hat{P}=\sum_{\boldsymbol{x}}P(\boldsymbol{x})\hat{I}_{\boldsymbol{x}}$, where $\sum_{\boldsymbol{x}}P(\boldsymbol{x})=1$. 
Consider now a trajectory generated by the dynamics starting from the point $\boldsymbol{x}_0$ in the configuratio space. Since $\mathcal{C}$ is finite and the dynamics are deterministic and invertible (i.e., a permutation), the trajectory corresponds to a closed orbit $\gamma_{\boldsymbol{x}_0}=\{U^t(\boldsymbol{x}_0)\}_{t=0}^{T-1}$ with a (minimal) period $T=|\gamma_{\boldsymbol{x}_0}|$, in which every configuration appears only once. Over such an orbit one naturally defines a flat probability distribution
\begin{align}
     \hat{P}^{(\boldsymbol{x}_0)}=T^{-1}\sum_{\boldsymbol{x}\in\gamma_{\boldsymbol{x}_0}} \hat{I}_{\boldsymbol{x}},
\end{align}
as well as its marginalization with respect to some finite contiguous subsystem $\Lambda$ of the lattice,
\begin{align}
\begin{aligned}
    &\hat{P}^{(\boldsymbol{x}_0)}_\Lambda=
    \sum_{\boldsymbol{x}_\Lambda} P^{(\boldsymbol{x}_0)}_\Lambda(\boldsymbol{x}_\Lambda)
    \hat{I}_{\boldsymbol{x}_\Lambda}, 
    \\
    &P^{(\boldsymbol{x}_0)}_\Lambda(\boldsymbol{x}_\Lambda)=
    \sum_{
    \mathclap{
    \substack{
    \boldsymbol{x}_{\bar{\Lambda}}: \\ \boldsymbol{x}_\Lambda\cup\boldsymbol{x}_{\bar{\Lambda}}\in
    \gamma_{\boldsymbol{x}_0}
    }
    }}T^{-1}.
\end{aligned}
\end{align}
Here, e.g., $\boldsymbol{x}_\Lambda=(\rho_1,\sigma_1,\rho_2,\sigma_2,\ldots,
\rho_{|\Lambda|},\sigma_{|\Lambda|})$ and $\bar{\Lambda}=\{|\Lambda|+1,\ldots,2L\}$ is the complement of the subsystem. For the local ergodicity to hold, the reduced probability distribution $\hat{P}^{(\boldsymbol{x}_0)}_\Lambda$ should converge to a (reduced) microcanonical distribution, defined on the subsystem $\Lambda$ as $\hat{P}^{\rm(mc)}_\Lambda=3^{-2|\Lambda|}\sum_{\boldsymbol{x}_\Lambda} \hat{I}_{\boldsymbol{x}_\Lambda}$ (note that the sum runs over $2|\Lambda|$ parameters). This convergence can be quantified using the subsystem-restricted $L^1$ distance
\begin{align}
    d_\Lambda(\gamma_{\boldsymbol{x}_0})\!=\!
    \|\hat{P}^{(\boldsymbol{x}_0)}_{\Lambda}\!-\!\hat{P}^{\rm(mc)}_\Lambda\|
    \!=\!
    \sum_{\boldsymbol{x}_\Lambda}
    \!\left| P^{(\boldsymbol{x}_0)}_\Lambda(\boldsymbol{x}_\Lambda) \!-\!3^{-2|\Lambda|}\right|.
    \label{eq:average-mean-distance}
\end{align}
Averaging over different orbits according to their configuration-space fractions, $T_\gamma/|\mathcal{C}|$, one obtains the \emph{average mean distance} $d_\Lambda=\mathbb{E}_{\gamma}[d_\Lambda(\gamma)]$. 
In Ref.~\cite{Orlov_2026}, $d_\Lambda$ was found to vanish as $d_\Lambda\sim e^{-\alpha L}$ ($\alpha>0$) in a (generic) cellular automaton without local conservation laws, while an algebraic decay $d_\Lambda\sim L^{-\beta}$ ($\beta>0$) was observed in the presence of (at least one) unresolved extensive conservation law.

\begin{figure}[t!]
\centering
\includegraphics[width=\linewidth]{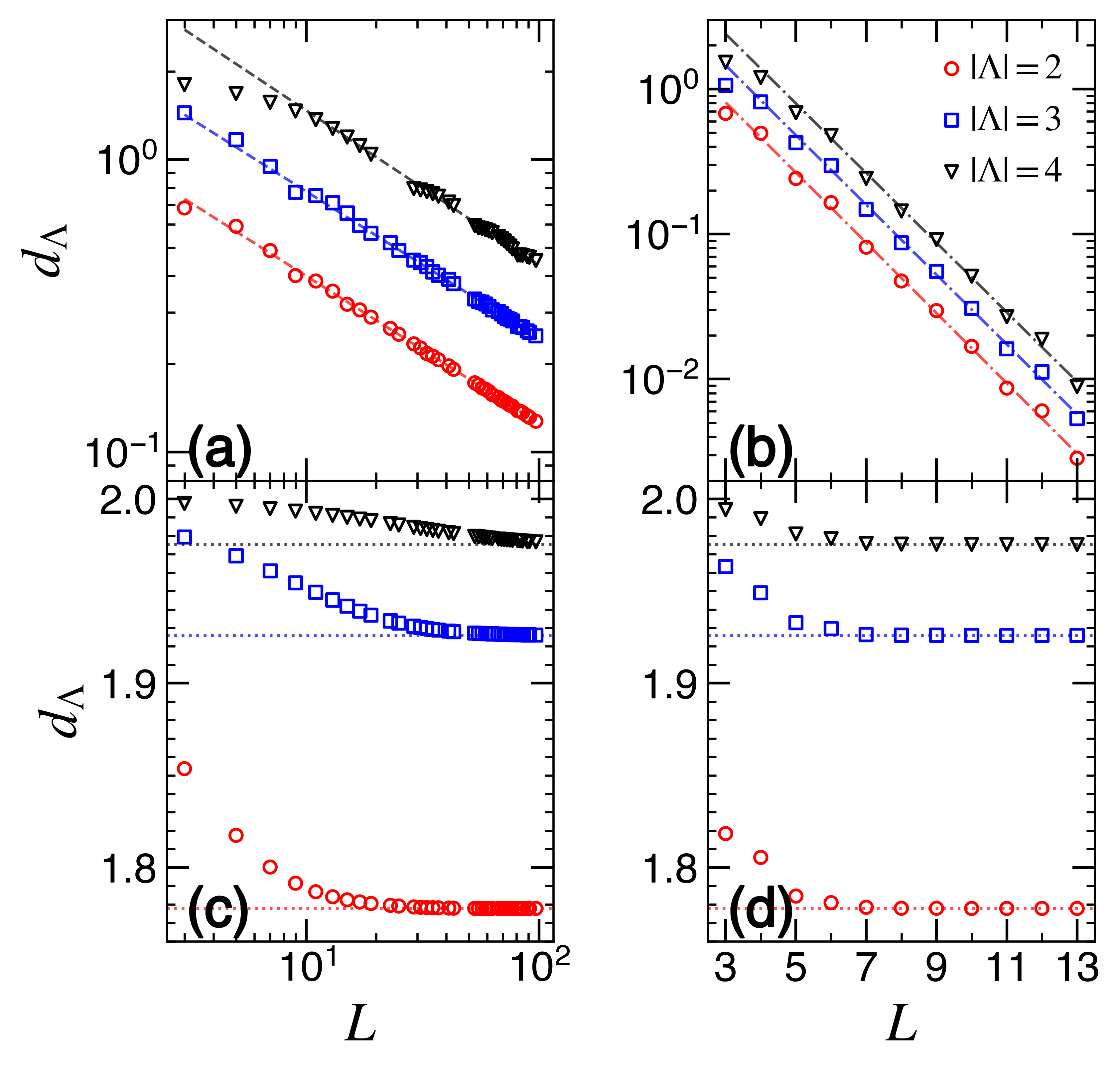}
\caption{\textbf{Ergodicity breaking in fragmented cellular automata.} Plotted are the average mean distances $d_\Lambda$ for automata with linear and nonlinear update rules, Eqs.~\eqref{eq:cellular_automaton} and \eqref{eq:nonlin-automaton} respectively. Panel (a) is for the linear model with resolved fragmentation. $d_\Lambda$ is evaluated within a single randomly chosen fragment and exhibits power-law decay, $d_\Lambda(L)\sim L^{-\beta}$, with fitted exponent $\beta=0.503\pm0.004$ (dashed lines). Panel (b) corresponds to the nonlinear model with resolved fragmentation. $d_\Lambda$ decays exponentially, $d_\Lambda(L)\sim e^{-\alpha L}$, with fitted $\alpha=0.554\pm0.020$ (dashed-dotted lines). Panels (c) and (d) are, respectively, for the linear and nonlinear models without resolved fragmentation. In both cases $d_\Lambda$ is evaluated over the full configuration space and saturates to a nonzero value $\bar{d}_\Lambda = 2(1-3^{-|\Lambda|})$ (dotted lines), defined in Eq.~\eqref{eq:saturation-distance}.} 
\label{fig:automaton}
\end{figure}

To check whether these observations hold in our automata, we study the scaling of the average mean distance after resolving the constraints. This amounts to fixing the parameters $\{\sigma_n\}$ and considering the dynamics in terms of variables $\{\rho_n\}$ only. In particular, it means that the full subsystem-reduced microcanonical distribution $\hat{P}^{\rm(mc)}_\Lambda$ in Eq.~\eqref{eq:average-mean-distance} should be replaced with the one defined on a fragment containing the initial configuration $\boldsymbol{x}_0=(\rho^\star_1,\sigma^\star_1,\ldots,\rho^\star_{2L},\sigma^\star_{2L})$, specifically,
\begin{align}
\begin{aligned}
    &\hat{P}^{(\boldsymbol{x}_0,{\rm mc})}_{\Lambda}=\sum_{\boldsymbol{x}_\Lambda}P^{(\boldsymbol{x}_0,\rm mc)}_\Lambda(\boldsymbol{x}_\Lambda) \hat{I}_{\boldsymbol{x}_\Lambda},
    \\
    &P^{(\boldsymbol{x}_0,\rm mc)}_\Lambda (\boldsymbol{x}_\Lambda) = 
    \begin{cases}
        3^{-|\Lambda|} & \text{if } \sigma_n = \sigma^\star_n \text{ for } n\in\Lambda,
        \\
        0 & \text{otherwise.}
\end{cases}
\label{eq:fragment-mc-distribution}
\end{aligned}
\end{align}
As shown in Figs.~\hyperref[fig:automaton]{4(a--b)}, our linear and nonlinear automata indeed exhibit the expected algebraic and exponential decay of $d_\Lambda$, respectively.

The picture changes if the constraints are not resolved: in this case the average mean distance, as a function of $L$, saturates to a nonzero value for both automata, as shown in Figs.~\hyperref[fig:automaton]{4(c--d)}. We find that the saturation value matches the distance
\begin{align}
    \bar{d}_\Lambda\!=\!\left\|\hat{P}^{(\boldsymbol{x}_0,{\rm mc})}_\Lambda\!-\!\hat{P}^{\rm(mc)}_\Lambda\right\|
    \!=\!\sum_{\boldsymbol{x}_\Lambda}\!\left|P^{(\boldsymbol{x}_0,\rm mc)}_\Lambda\!(\boldsymbol{x}_\Lambda)\!-\!3^{-2|\Lambda|}\right|
    \label{eq:saturation-distance}
\end{align}
between the subsystem-reduced microcanonical distributions with and without fixed constraints, respectively, which can be computed exactly. Specifically, the sum in Eq.~\eqref{eq:saturation-distance} runs over $3^{2|\Lambda|}$ subconfigurations $\boldsymbol{x}_{\Lambda}$, out of which only $3^{|\Lambda|}$ have $\sigma_n=\sigma^\star_n$ for $n \in \Lambda$. Each one of these contributes $3^{-|\Lambda|}-3^{-2|\Lambda|}$ to the sum, while the rest of the terms each contribute $3^{-2|\Lambda|}$, together yielding $\bar{d}_\Lambda=2(1-3^{-|\Lambda|})$.
The result confirms the expectation that the orbit $\gamma_{\boldsymbol{x}_0}$ ergodically explores only the fragment of the full configuration space that contains the initial configuration $\boldsymbol{x}_0$. 

\section{Discussion}

We studied quantum and classical many-body lattice dynamical systems with an extensive number of localized constraints that partition the Hilbert or configuration space into dynamically disconnected sectors, or the classical Hamiltonian phase space into invariant submanifolds. For the quantum model and the classical systems with discrete degrees of freedom, this splitting is uniform: the sectors are exponentially numerous and equal in size. Our main motivation was to investigate the consequences of this simple form of fragmentation for transport and thermalization.

Our first example was the quantum Volterra model, a dynamical system closely related to integrable lattice discretizations of certain field theories. By resolving a maximal commuting subset of the localized constraints, we mapped the dynamics within each Hilbert-space fragment onto a quantum chiral Potts chain. Taking a nonintegrable classical limit of the quantum Volterra model, we then showed that the localized constraints do not generate a lattice gauge symmetry. We subsequently studied energy transport in both the quantum model and its nonintegrable classical counterpart, finding that the localized constraints do not, by themselves, give rise to anomalous transport. In the quantum model, the Drude weight, which characterizes ballistic transport, is identical across all Hilbert-space fragments and is consistent with that of the full system, in which the constraints remain unresolved. Similarly, in the nonintegrable classical limit, the diffusion constant is independent of the values taken by the constraints and coincides with that computed over the full phase space.

The effect of localized constraints is more directly visible in ergodic properties. We investigated this in classical cellular automata derived from the constrained dynamics of the nonintegrable classical limit. In these fully discrete systems with a finite configuration space, we demonstrated that local ergodicity breaks down when the constraints remain unresolved.

Taken together, our results show that, while localized constraints strongly restrict the accessible state space and thereby obstruct ergodicity, their consequences for transport are more subtle. In the fragmented systems that we considered, the decomposition into dynamically disconnected sectors does not by itself imply anomalous transport. Instead, transport properties are controlled by the dynamics within individual sectors and, when these differ, by the interplay of their contributions.

\textit{Note added.}---Shortly after posting this work, we became aware of two independent works~\cite{Pushkar_2026,Hamazaki_2026} investigating the effects of noncommuting local conserved quantities on ergodicity. Ref.~\cite{Hamazaki_2026}, which appeared concurrently with our work, identifies a closely related form of fragmentation termed \emph{Hilbert-subspace cloning}.

\section*{Acknowledgements}

L.Z. would like to thank E.~Ilievski, O.~Motrunich, and V.~Korepin for valuable discussions.
This work has been supported by the Slovenian Research and Innovation Agency (ARIS) through the University of Ljubljana Startup Project NESY (contract No. SN-ZRD/22-27/0510), Research Projects Nos. N1-0504 and N1-0368 (L.Z.), and Research Program No. P1-0402 (L.Z.). L.Z. also acknowledges support from the European Research Council (ERC) through Advanced Grant QUEST (Grant Agreement No. 101096208).

\bibliography{references.bib}

\end{document}